# What's happened in MOOC Posts Analysis, Knowledge Tracing and Peer Feedbacks? A Review

Manikandan Ravikiran

mravikiran3@gatech.edu

## 1 INTRODUCTION

Learning Management Systems (LMS) and Educational Data Mining (EDM) are two important parts of online educational environment with the former being *a centralized web-based information systems where the learning content is managed and learning activities are organized* (Stone and Zheng, 2014) and latter focusing on using data mining techniques for the analysis of data so generated. As part of this work, we present a literature review of three major tasks of EDM (See section 2), by identifying shortcomings and existing open problems, and a *Blumenfield* chart (See section 3). The consolidated set of papers and resources so used are released in `https://github.com/manikandan-ravikiran/cs6460-Survey`. The coverage statistics and review matrix of the survey are as shown in Figure 1 & Table 1 respectively. Acronym expansions are added in the Appendix Section **??**.

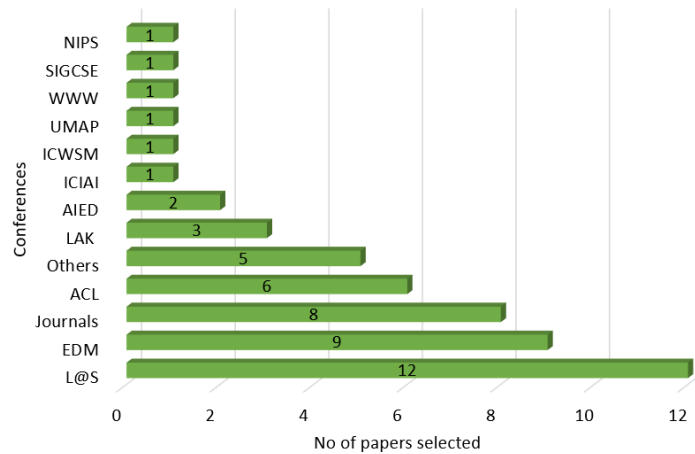

*Figure 1*—Coverage statistics of papers used in survey. *ACL:= NLP-BEA Workshop, Journals:= (Computer Education, ACL Transactions, Intelligent Tutoring Systems, User Modeling and User Adapted Interaction, Neurocomputing), Others:= (Arxiv)*

*Table 1*—Literature review matrix of selected papers. P:= Precision, R:=Recall, F:= F1 Score, CV:= Cross Validation, T/T:= Train Test Split, ROC AUC:= Receiver operating characteristics Area Under Curve, Stanford:= Stanford MOOC Post Corpus, * := Unclear, LDS:= Lingusitc/Data Specific Feature, RD:= Raw Data, SPP:= Student Performance Prediction

| Reference | Outcome | Data Source | Algorithms | Features | Performance Metrics | Prediction Architecture |
|---|---|---|---|---|---|---|
| (Akshay et al., 2015) | Confusion Prediction | Stanford | Logistic Regression | BOW, Meta data | P, R, F1 | T/T |
| (Bita et al., 2018) | Student Assessment | Internal | Logistic Regression, LSTM | Student interactions | P, R, F1 | T/T |
| (Alam et al., 2018) | Categorization | Internal | Random Forest Importance; Chi Square Artificial Neural Network | Data Specific | P, R, F1 | T/T |
| (Omaima, Aditya, and Huzefa, 2018) | Urgency Classification | Internal | Adaboost | LDS | P, R, F1 | T/T |
| (Jaime and Kyle, 2015) | Speech Act Classification | Internal | Logistic Regression | LDS | AP | CV |
| (Bakharia, 2016) | Post Classification | Stanford | Adaboost, Random Forest, SVM | LDS | P, R, F1 | CV |
| (Sumit, Chuck, and Lucy, 2013) | SAG | USCIS | Logistic Regression | LSA, TFIDF | ROC AUC | T/T |
| (Steven et al., 2012) | Essay Scoring | Internal | COGENT | LSA, TFIDF | Coverage and Retrieval | T/T |
| (Brooks et al., 2014) | SAG | Internal | K medoids | LSA, TFIDF | Custom | T/T |
| (Scott et al., 2015) | Knowledge Tracing | Internal Posts | WAT, TAALS, TAAS | LDS | P, R, F1 | T/T and CV |
| (Farra, Somasundaran, and Burstein, 2015) | Essay Scoring | TOEFL Essays | Logistic Regression | LDS | P, R, F1 | T/T |
| (Guo et al., 2015) | Performance Prediction | Internal | Deep Learning | RD | Coverage and Retrieval | T/T |
| (Harrak et al., 2019) | Post Classification | Internal | Multiple | LDS | Cohen Kappa(κ) | CV |
| (Kakkonen et al., 2005) | Essay Scoring | Internal | Latent Semantic Analysis | * | Coverage and Retrieval | T/T |
| (Mohammad et al., 2014) | Knowledge Tracing | Internal | Item Response Theory - FAST | * | ROC AUC | T/T |
| (Kim, Vizitei, and Ganapathi, 2018) | Performance Prediction | Internal | Deep Learning | RD | ROC AUC | CV |
| (Kuzi et al., 2019) | Assignment Assisment | Internal | probabilistic topic model | Topical Features | Kendall's Tau | T/T |
| (Lalwani and Agrawal, 2017) | Knowledge Tracing | Internal | Deep Learning | Knowledge Graph | ROC AUC | T/T |
| (Liu et al., 2016) | Sentiment | Internal | Swarm Optimization | LDS | P, R, F1 | T/T |
| (Madnani et al., 2013) | Essay Scoring | Internal | Scoring | LDS | Coverage and Retrieval | T/T |
| (Sharada, Shashi, and Xiong, 2018) | Knowledge Tracing | ASSistments | Deep Learning | RD | ROC AUC | T/T |
| (Nagatani et al., 2019) | Knowledge Retention | ASSistments | Deep Learning, Random Forest | RD | R Squared | T/T |
| (Northcutt, Leon, and Chen, 2017) | Knowledge Tracing | ASSistments | Deep Learning | RD | ROC AUC | T/T |
| (Okubo et al., 2017) | SPP | Internal | Deep Learning | RD | AP | T/T |
| (Pandey and Karypis, 2019) | Knowledge Tracing | ASSistments | Deep Learning with Attention | RD | ROC AUC | T/T |
| (Philip, Hao, and Kenneth, 2009) | Knowledge Tracing | Internal | Factor Analysis | RD | BIC, Liklihood | T/T |
| (Piech et al., 2015) | Knowledge Tracing | ASSistments | LSTM | RD | ROC AUC | T/T |
| (Ramesh et al., 2014) | MOOC discussion | Internal | Seeded LDA | Topical Features | ROC AUC | T/T |
| (Andres et al., 2018) | SPP | Internal | Multiple | LDS | P, R, F1 | T/T |
| (Taghipour and Ng, 2016) | Essay Scoring | ASAP | Deep Learning | RD | Cohen Kappa(κ) | T/T |
| (Tang, Peterson, and Pardos, 2019) | Sequential Data tasks | Internal | Deep Learning | RD | * | T/T |
| (Tucker, Dickens, and Divinsky, 2014) | Sentiment | Internal | Internal | LDS | P, R, F1 | T/T |
| (Waters, Tinapple, and Baraniuk, 2015) | Peer Grading | Internal | Bayesian Ranking | LDS | Rank-1 | * |
| (Yang et al., 2015) | Confusion Prediction | Internal | Logistic Regression | LDS | ROC AUC | T/T |
| (Yeung and Yeung, 2018) | Knowledge Tracing | ASSistments | Deep learning | RD | ROC AUC | T/T |
| (Yin, Moghadam, and Fox, 2015) | Peer Grading | Internal | Clustering | LDS | ROC AUC | * |
| (Zhang, Shah, and Chi, 2016) | ASG | Internal | Multiple | LDS | P, R, F1 | T/T |
| (Sales et al., 2018) | ASG | ASAP | Memory Models | Glove | QWK Scores | T/T |

## 2 LITERATURE REVIEW

The literature review is divided into three parts namely reviews of **MOOC Post analysis**, **Knowledge Tracing** and **Peer Feedbacks**. Further, each review accommodates a meta-analysis and open problems that could be further explored and addressed.

### 2.1 MOOC Comments and Post Classification

Over the past two decades, much of the works on EDM concerning MOOC Post Classification (MPC) a.k.a discussion forum post text classification have focused majorly on structured datasets. The works vary according to their final intended goal itself, resulting in a diverse range of datasets, features, and algorithms. In this review, we analyze MPC based on diversity of end goals i.e. *classification of posts for XYZ task*. Table 1 shows various datasets/approaches used as part of the important works.

**Student Intervention:** The earliest works in these focuses on confusion prediction using a classification approach on clickstream datasets (Yang et al., 2015). The aim was to *improve the identification and intervention of the student's confusion* as part of MOOC auditing. The work uses linguistic features, question features and clicks patterns features and analyses logistic regression model with 10 fold cross-validation for survival curves across the various courses. It finds that the more students express their confusion and are exposed to confusion in the MOOC forums, the less likely students are to remain active in the learning community and helping to resolve or providing responses to student confusion reduces their dropout in the courses (Yang et al., 2015). Further, it shows that the extent to which different confusion affect dropout is determined by specific courses, emphasizing the need for the cross-modal analysis.

**Student Success, Completion and Retention:** Much of MOOC's and online courses generate large amounts of textual data in recent times. As such application of NLP in EDM for predicting success has seen large traction. Early works include that of Scott et al. (2015) which *analyzed three tools namely Tool for the Automatic Analysis of Lexical Sophistication (TAALES), Writing Assessment Tool (WAT), & Tool for the Automatic Assessment of Sentiment (TAAS),* to understand lexical complexity, identification to post length, situation cohesion, cardinal numbers, quality of writing, and tri-gram/bi-gram frequency etc. The results showed that when pre-dicting MOOC completion, considering the writing quality is more useful than assessing observed behaviors.

Then there are works by Ramesh et al. (2014) which included linguistic and behavioral features of MOOC discussion forums to predict student survival using seeded latent Dirichlet allocation showing linguistic analysis of user posts. The main goal of the work was *analysis of MOOC content for improving student retention and helping in starting instructor intervention*. It analyzed coarse-grained, grouping posts into three categories where all the meta-content, course logistics, and course feedback were grouped under the same topic category. The results suggest that two students survive the course and a timely answer to their posts might have been a reason influencing these students to complete the course. However, such an analysis at a deeper level is missing.

**Sentiment Prediction:** Besides above, there are works of Liu et al. (2016) and Tucker, Dickens, and Divinsky (2014) which focus on predicting sentiments, with the former presenting a novel feature selection method for *sentiment recognition reviews with a motivation of reducing the dimensionality and redundancy of feature space* to get an F-Score of 88%. The work by Liu et al. (2016) used the sentiment to help educators to understand the factors that may impact student performance, team interactions, and overall learning outcomes using swarm optimization. The authors analyze the developed system on an actual MOOC course study and show the benefit of an analysis of posts by students. Finally, there is the work of Wen, Yang, and Rosé (2014), which shows the *usefulness of the sentiment results by emphasizing a strong correlation to drop out behavior*. While much of these works are dissimilar in the dataset used and their end aim of the analysis, the core tech is fairly similar (Text-based modeling).

**Confusion Prediction and Recommendation:** On the parallel side, some works use posts and their metadata to detect confusion in the educational contents. Notable work by Akshay et al. (2015), *emphasizes the capacity of posts to improve content creation*. The work uses the Stanford MOOC posts dataset and presents an analysis of six of its subsets along with their videos where the authors further test the relevance of the recommendation based on data from the posts. The core aim of the work is in investigating & strengthening video snippet ranking by considering which video portions learners re-visited. The work overall presents high accuracy on the multiple datasets. Besides the previously mentioned works on analysis of sentiments, confusion prediction, dropout predictions there are addi-tional works which presents the usefulness of predicting the success of students in MOOC using their discussion posts (Robinson et al., 2016).

**Improving Recommendations and Cross-Domain Analysis:** Discussion posts implicitly capture context and behavior of the student. In recent times, the diversity of applications on MPC has grown considerably by drawing parallel ideas from areas of Natural Language Processing (NLP) and Computer Vision (CV) to exploit such contexts. On one side we can see the works of Northcutt, Leon, and Chen (2017) which focuses on *Gated Convolutional Neural Networks and Maximal Marginal Relevance (MMR) based re-ranking with intention comment diversification* in the discussion forum's helping students to get a broader view of the posts. While the work shows diversification in comments, much of the work is plagued by unclear parameter selection, lack of relationship to hypothesis and experiments, small data sample and ill-defined evaluation metric. Moreover the end objective of diversification is unclear.

Besides ranking, there are works on cross-domain post classification, *with a focus on the analysis of the domain shift of developed a model (CNN-LSTM)* (Bakharia, 2016). The author through their experiments highlight the needs of transfer learning and domain adaptation algorithms and also provides insights into the algorithms required within an educational context. However, the work emphasizes on low cross-domain classification performance, warranting more works.

**Scaling Coding (Annotation) Schemes:** If building new models for some prediction task is one end, while the other end researchers also focus on developing and improving annotations of the posts itself. In this side, works by Harrak et al. (2019) is famous, which extended the pre-defined coding scheme of MPC to understand performance predictions. More specifically *the work focused on the extension of the coding scheme, first to understand the reliability of annotation of questions extracted from MOOC forum posts according to a fine-grained multi-level coding scheme and its usage to analyze the relationship between students' questions and their performance in the MOOC* (Harrak et al., 2019). While the work explicitly identifies the latter as the result, the former is ill-defined as the work uses Cohen's kappa($\kappa$) measure between the manual annotation and an automatic annotator to emphasize on accuracy, which has no significance, rather high inter-annotator agreement was warranted. Further, the experimentation with the coding scheme is too shallow because of the very small sample size Harrak et al. (2019). Also, because of lack of inter-annotator agreement and bias in splits, the work requires meticulous re-analysis.

**Others:** Finally, there are two major recent works in MPC. First, of which is the identification of urgency of posts (Omaima, Aditya, and Huzefa, 2018) where the goal *is to bring order to the chaos in MOOC discussion by classification of posts into content vs. non-content related posts* with multiple granularities of dataset splits. The work uses metadata and term frequency information using multiple classifiers, where it shows the ability to use a few linguistic features with few metadata to build a moderately to substantially reliable classification model that can identify urgent posts in Stanford MOOC discussion forums using 10-fold cross-validation achieving near 80% F score. Second, there is the work of speech act prediction from forum posts by Jaime and Kyle (2015) *to identify that combining redundant speech act labels from crowdsourced workers can approximate the labels from an expert and investigated the usefulness of speech acts for predicting instructor intervention, assignment completion, and assignment performance* (Jaime and Kyle, 2015). While speech acts were the most useful for predicting instructor intervention, the authors concluded that the speech acts were not as useful for predicting assignment completion and performance.

### 2.1.1 *Meta Analysis:*

1. **Feature & Algorithms:** From the survey, one can see three important facets of research. First, many of the work so produced uses a simple feature-based learning approach, with a few of the tasks recently developed using both traditional and deep learning approaches. However, given the empirical nature of the development process of models, there is no one-size-fits-all solution to set the best configuration for a specific problem and depends on the input data available and the task at hand. Among those analyzed, the biggest issue is that the depth of details on the parameter selection & unclear data selection. So are the intuitions behind the selection of features for algorithms.
2. **Benchmark and Datasets:** The MPC tasks are sparse (multiple varieties) and are plagued by a series of problems beginning with a lack of availability of the dataset itself. The main problem across the tasks is that there is not a single "correct" benchmark that can be assessed and this is because most of the datasets are very specific to individual universities and their courses. Because of such limited scope, there is already a *crisis of replication* (Andres et al., 2018).
3. **Annotation Guidelines and Evaluation Metrics:** MOOC classification problems possess a wide variety of annotation guidelines some of which are problem-specific and rest are related to implied application usage themselves. Majority of the works, which tackles a new problem and new annotation either use crowdsourcing or internal annotation process, but lack details on the inter-annotator agreement, annotator selection process, data separation mechanism and finally the evaluation metric itself a big problem which requires revisiting.

4. **Relationship between the problem and experimentation:** All the works presents a well-defined problems and a set of basic experiments to show the performance. However more depth and breadth in analysis is needed.

### 2.1.2 *Open Problems:*

1. **Feature & Algorithms:** Besides intuition behind the selection process for features, algorithm and parameter selection problems. The biggest of them being lack of use of language modeling strategies and active learning based ideas. Especially, with the scarcity of datasets. Ideas like transfer learning on BERT (Devlin et al., 2019) and RoBERta (Liu et al., 2019) are ideal fit for low resource problems of MPC's. The number of works till date are fairly limited (Veeramachaneni, O'Reilly, and Taylor, 2014).

2. **Benchmark and Datasets:** With varying datasets producing diverse results and with multiple new open problems on MPC. There is a need for a unified benchmark framework which provides a collection of well defined resources for training, testing, and analyzing the results.

3. **Relationship between the problem and experimentation:** There is need of detailed ablation studies under a single unified framework which probes the model on incomplete input, shuffle inputs, random labeling and random content replacement. All these scientifically ground the validity of the developed work.

## 2.2 Knowledge Tracing

Knowledge Tracing (KT) is a task of predicting student performance based on their activities, where the activities are represented by a sequence of variables. Unlike MPC, the task is unique, hence much of the work focuses on accuracy and interpretability rather than the end goal. In this section, we present the literature based on the approaches and datasets used to solve KT.

**Bayesian Knowledge Tracing:** KT has seen a significant amount of works over the past decade beginning with early works of Bayesian Knowledge Tracing (BKT) (Albert and John, 1994) which models students activity as a set of binary variable which represents understanding/not understanding of a concept. The work used an Hidden Markov Model (HMM) to update the probabilities across these defined variables, as a learner solves the concept correctly or incorrectly. The work assumed lifelong remembering behaviour in humans, which was a major flaw, as humans forget the concept after some period. Multiple other variants of BKT work followed this and dominated the area for over a decade notable among them includes Partially Observable Markov Decision Processes (Rafferty et al., 2011), Performance Factors Analysis (PFA) framework (Philip, Hao, and Kenneth, 2009) and Learning Factors Analysis (LFA) framework (Hao, Kenneth, and Brian, 2006) and Item Response Theory (IRT) models with switched nonlinear Kalman filters (Mohammad et al., 2014). Each of these models are extensions of original HMM addressing its inherent issues.

**Deep Knowledge Tracing:** While the Bayesian approaches governed the area over a decade, the controversial paper of Deep Knowledge Tracing (DKT) (Piech et al., 2015) was introduced in 2015 which modeled KT using Deep Recurrent Neural Networks. It dominates much of the KT by deep learning based approaches of RNN's and LSTM's, as it produces 25% improvement over earlier BKT models. The original DKT work tested the results in deep across multiple datasets ranging from simulated data, Khan Academy Data, to the Assistments benchmark dataset, with comprehensive analysis. Yet it had a fair amount of problems in the works, which is explained in the next section.

**Empirical Reviews:** Following DKT, there were a series of works that focused on deeper analysis, questioning the results of Deep Knowledge Tracing (DKT). There is set of works which presents issues in DKT notable works include that of (Lalwani and Agrawal, 2017), (Mao, Lin, and Chi, 2018) and (Wilson and Xiong, 2016) and those which agree and highlights the importance of the overall results (Lin and Chi, 2017), (Wang et al., 2017) and (Montero et al., 2018).

Notable among the DKT favouring works is that of Lin and Chi (2017) who compared a series of Bayesian Knowledge Tracing (BKT) models against vanilla RNNs and Long Short Term Memory (LSTM) based models. The results showed that the LSTM-based model achieved the highest accuracy, and the RNN based model had the highest F1-measure. It also found that RNN can achieve a reasonably accurate prediction of student final learning gains using only the first 40% of the entire training sequence, which is higher compared to that of BKT.

On the opposing side of DKT, work by Wilson and Xiong (2016) is famous, which points out an issue in the dataset so used and the experimental setting fairly comparing the entire controlled experiments on traditional ML algorithms arguing that the difference in deep learning and traditional ML as stated is not clear. Finally, there are some extensions to DKT which focuses on a different set of goals focusing on prediction of student pass or failing the assignments and their analysis. However, the works differ only in their aim.

**Dataset Types:** Further within KT there works that can be differentiated across the two sources of datasets so used for the task namely writing samples and clickstream datasets, with latter dominating the research. Notable works in this include that of Tang, Peterson, and Pardos (2019) which used both of the variants to highlight the application and flexibility especially in deep learning and sequential student data. It trained a two-layer Long Short-Term Memory (LSTM) network on two distinct forms of education datasets of essays and MOOC clickstream data to show the network attempts to learn the underlying structure of the input sequences. Further, they show that the so developed model is useful to produce new sequences with the same underlying patterns exhibited by the input distribution targeting low sample size issues in EDM. In related lines, there are also works that include the analysis of knowledge retention on data from web-based mathematics class using KT on sequential data (Sharada, Shashi, and Xiong, 2018).

Besides clickstreams and writing assignment texts, there are additional works in using log data. Interesting works include that of Okubo et al. (2017) that proposes a method for the prediction of grades based on LMS log data, e-portfolio, and e-book system logs. Besides prediction, there are works in the lower level of granularity especially that of Alam et al. (2018) that focus on categorizing the students into high, medium and low one exhibiting knowledge tracing.

**Unsupervised Research:** While supervision has dominated KT research over the years there are few which tried using unlabeled datasets. In this line, the interesting works include that of Guo et al. (2015) where the authors semiautomatically annotated the unlabeled data representations at multiple levels and highlighted the effectiveness of the developed method.

**Domain Adaptation:** From the perspective of domain adaptation, there is GritNet (Kim, Vizitei, and Ganapathi, 2018) that produces substantial gains by re-cast the student performance prediction problem as a sequential event prediction problem.

**Recent Works:** Much of the recent works has come in three directions namely adapting novel deep learning algorithms to the problem (Pandey and Karypis, 2019), improving overall performance through additional inputs through knowledge augmentation (Nagatani et al., 2019) and addressing fundamental problems after application of novel deep learning algorithms (Yeung and Yeung, 2018), involving reducing cost sensitivity via regularization.

### 2.2.1 *Meta Analysis:*

1. **Feature & Algorithms:** From the survey, we can see two major aspects first being that most works focus on establishing fairness across the experimentation between Bayesian and Deep models, resulting in much debate on the benefits/reliability of both. However, perhaps either of them are beneficial, where the former presents interpretability and the latter shows performance. Considering the status of active research in deep learning, one can expect more applications of deep learning methods.
2. **Benchmarks & Datasets:** Among all the areas of the EDM attempted in this assignment, the datasets and benchmarking are very consistent only in the case of the Knowledge Tracing tasks. There are a significant amount of datasets that are easily available for active research. Further, the area of KT is dense even in cross-domain datasets, especially with more or more works of prompting the need for significant research in cross-domain analysis and generalization.
3. **Annotation Guidelines & Evaluation Metrics:** Similar to benchmarks, the annotations are fairly consistent across the released datasets. However, experimental investigations by Wilson and Xiong (2016) who showed that in some large datasets multiple inputs were tagged with multiple skill labels. This caused the deep learning methods to perform well, as it processes all inputs simultaneously, providing the model access to ground truth when making a prediction. However, the metrics of evaluation are well-grounded.
4. **Relationship b/w problem and experimentation:** Much of the literature has two sides to experimentation with one side targeting improvement of results, while on the other side arguing on the sanity on the experimental finding from the other side. Both have led to conflicting results and analysis. However, in recent works of (Pandey and Karypis, 2019) & (Yeung and Yeung, 2018) we can consistently see that some of these sanity arguments are well addressed .

*2.2.2 Open Problems:*

1. **Feature & Algorithms:** While LSTM's have dominated the area of knowledge tracing, they do show problems of uncertainty, low cross-domain performance, etc. Approaches like LSTM that exploit current input and memory, usually lack generalization power across the domains. In this sense, attention-based approaches have a higher probability of showing performance higher across the domains. We can see this already happening through works of (Pandey and Karypis, 2019), but we need more works.
2. **Bench marking & Datasets:** Dataset has been both boon and a curse for this task, especially with works of (Wilson and Xiong, 2016) showing the issues in the datasets itself. This warrants significant study on major open large datasets under a common framework where underlying theory and relationship to the application are preserved. The net efforts on cross-domain analysis and development are still in a nascent stage warranting, more works to be produced.
3. **Annotation Guidelines & Evaluation Metrics:** Deep Knowledge Tracing, requires bias and fairness evaluations across all the released datasets, especially with the risk of duplicating the skill labels across the data rows. Looks like multiple works (Yeung and Yeung, 2018) post the paper of (Piech et al., 2015) still reporting results on the same duplicated datasets, which need to be revisited.
4. **Relationship between defined problem and experimentation:** While the experiments are well designed, a plurality of papers only project improvement and lack details on error analysis. An active research problem that's warranted includes replication of the experiments and presentation of errors, grounding the work fairly and scientifically.

## 2.3 Peer Feedback & Grading

Peer feedback, a task of viewing and critiquing others' work plays a key pedagogical role in MOOC's especially with the advent of more creative courses in recent time. In a general learning setting colocation of students provide a shared context for the students thereby conferring values to students' work. As such, scaling peer assessments including both evaluation and peer learning has been an important problem with numerous works among which providing evaluation tools to help TA's and peers is notable. The approaches so developed for this can be classified into three subtasks namely Automated Essay Scoring (AES), Au-tomatic Short Answer Grading (ASG) and other evaluation tools. Before review, it needs to be highlighted that AES has large amount of works from linguistics community as well, however much of our works looks from perspective of Edtech community with few from linguistics works.

**AES:** This task typically involves the evaluation and scoring of essays. This problem has seen a substantial collection of works, notable of them include that of Zhao et al. (2017) which proposed a memory augmented neural model. The primary intuition behind this is that grading samples for each score in the rubric, using such samples to grade future similar work. The model so developed learned to predict a score by computing the relevance between the students' responses and the grading criteria collected. The work is critical in experimentation with grades ranging from 7 to 10 to outperform and improve the average QWK score by 4% compared to the baseline LSTM on Kaggle Automated Student Assessment Prize (ASAP). Despite memory augmentation being like that of LSTM, it's interesting to see exploiting temporality of similar assignments leading to improvement in results (Similar to human cognition - More experience better results).

An alternative to memory networks comprises works of Taghipour and Ng (2016) which tested multiple neural network models for automated essay scoring and learn the best feature representation to learn the relation between an essay and its assigned score. Results showed an improvement of 6% over other approaches requiring feature engineering. Further, the approach presents details on the failure of attention networks, which are typically as known to perform well on language tasks. The work highlights the interpretability nature of the model so got and shows appropriate indicators of essay quality are being learned, including essay length and essay content. We can see additional works in similar lines by (Steven et al., 2012), (Madnani et al., 2013), (Kakkonen et al., 2005) and (Farra, Somasundaran, and Burstein, 2015) respectively. We list more papers in https://tinyurl.com/wzo8ybg.

**ASG:** Similar to AES, ASG has seen a fair bit of traction, especially where ASG systems typically focus on automatically classify students' answers as correct or not, based on a previous set of correct answers. Notable works, on ASG, includes that of Zhang, Shah, and Chi (2016) which studied feature from the answer, questions, and student models, both individually and combined, integrating them in different machine learning models. The work highlights that the deep learning model got the best performance in their experiments. In line with this, the work compared several features for the classification of short open-ended answers, such as n-gram models, entity mentions and entity embeddings. The authors got inconclusive results regarding the benefits of using embeddings regarding traditional n-grams. Multiple works on ASG have focused on topic modeling including that of Kuzi et al. (2019) which proposes to study multiple topical modeling features and combining them with simple lexical features and verifies the same on clinical case assignments to show effectiveness and improvement. However, it can be seen that topic modeling is very sensitive to topics themselves producing an array of results.

An alternative to straight classification, ASG has seen a fair bit of clustering works especially focusing on understanding a coarse-grained view of the submitted assignment and simplification of peer reviews. There are two major works first of which is focuses on multiplication of the instructor's leverage (Yin, Moghadam, and Fox, 2015), by grouping student submissions according to the general problem-solving strategy so used. The work finds that it is possible to automatically create clusters such that an instructor eyeballing some representative submissions from each cluster can readily describe qualitatively what the common elements are in student submissions in that cluster. Second in this line of work include (Brooks et al., 2014) which proposed a cluster-based interface that allows teachers to read, grade, and provide feedback on large groups of answers at once. The work experimented with the developed system with 25 teachers resulting in teachers grading far more quickly using the clustered version, and that the resulting grades being like that of the gold standard used for evaluation.

**Others:** The categories of works other than AES and ASG are fairly broad focusing on multiple nuances of grading and peer assessment. For example, works of Bita et al. (2018) use a temporal analytics framework for stealth assessment of student problem-solving strategies in a gamed based learning environment by modeling problem-solving behaviors. Alternatively, there are works that explored how a DL-based text analysis tool could help assess a student's moral thinking (Heeryung et al., 2017). However, the work lacked deeper experimentation. Alternatively, there are assistant tools including that of Sales et al. (2018) which also proposed a method to estimate mastery of skills using A/B tests. This paper provides theoretical conditions for improving statistical precision and estimates of various effects. In the same category of works we have coding com-position emulators notable of them being Automated Coding Composition Evaluator (Rogers, Tang, and Garcia, 2014) which automated grading of programs by assessing composition of code through static analysis, conversion from code to AST, and clustering (unsupervised learning), helping automate the subjective process of grading based on style and identifying common mistakes.

Finally, there are works on predicting students' performance using assignments with notable works namely by Sánchez-Santillán et al. (2016) focusing on building more accurate classification models to predict the output, analyzing the interaction incrementally. The work studied the temporally gathered data on and studied if it is possible to get more accurate classification models through the analysis of the students' interaction incrementally, rather than doing it all at once to show that it is possible to get better classification models using an incremental interaction. Multiple more works exist ranging from Bayesian re-ranking (Waters, Tinapple, and Baraniuk, 2015) to improve peer grading, power grading approaches using cluster analysis (Sumit, Chuck, and Lucy, 2013) to optimize the ASG process and works that focus on peer assessment in low resource setting (Molapo et al., 2019).

### 2.3.1 *Meta Analysis:*

1. **Feature Algorithms:** From the survey, we can see that we have covered the two problems of ASG and AES with a wide variety of algorithms. With AES, we can attribute this to their interest in the linguistic community. We can see that the range of features is diverse some derived from data and some from topic modeling and so on. I can see similar behavior even with ASG. Added to this ASG has seen numerous works from unsupervised approaches.

2. **Benchmark Datasets:** For AES, Benchmark and datasets have two sides to it. During our survey, we find that papers that are submitted to Ed-tech conferences like L@S, EDM, etc. present results on a wide variety of benchmarks, but those from computational linguistic communities are always on similar benchmarks. This is because of the end goal itself, where the former focuses on the usage of AES in end applications and the latter focuses on algorithmic improvements. As such behavior of benchmarks and datasets is very similar to that of the previous both MPC and KT. However, for ASG the findings are consistent with that of the MPC task.

3. **Annotation Guidelines and Evaluation Metrics:** The annotation guidelines are fairly easier in both AES and ASG, so are evaluation metrics which mostly restricts to Cohen's Kappa (κ) measure.

4. **Relationship between problem and experimentation:** The experimentation across all papers fairly supports the results and addresses the problems.

### 2.3.2 *Open Problems:*

1. **Feature & Algorithms:** AES and ASG because of its traction in linguistic communities have seen significant new works with novel approaches that are designed in data mining and machine learning. However, unsupervised AES and active learning in AES is something still in a nascent stage.
2. **Benchmark & Datasets:** From this survey, we find that many benchmarks and datasets for cross-domain AES and ASG are fairly limited. However, this requires further literature exploration.
3. **Annotation Guidelines & Evaluation Metrics:** Multiple works of AES relies on a single score for the entire work and this is true across both the segments of linguistics and EDM communities. However, it is generally seen that there is a fixed rubric for scoring and generally peer's address according to this. *A work in this line would be to find intersections of MPC and AES to build better TA/Peer assisting tools*.
4. **Relationship between problem & experimentation:** While the experimentation sufficiently backs up the problem description, the addition of in-depth error analysis of the model's criteria for score prediction would be more useful across all works. Also to date, the works are fairly limited on the impact of the rubric in peer grading, which is one avenue for exploration.

## 2.4 Discussion

In this work, we surveyed three major tasks of EDM namely MPC, KT and Peer Feedbacks (AES & ASG). We reviewed a total of 51 papers, obtained based on a systematic search on the three mentioned problems. During our survey, we found multiple dimensions works under each tackling application-centric, algorithmic and data specific challenges through improved annotations. At the same time, we saw multiple open problems and challenges starting from usage of simple data-driven techniques, the crisis of replication in case of MPC due to issues with datasets, lack of certainty in results from KT and limited cross-domain works in case of AES and ASG. At the same time, we found consistency in results and evaluation metrics across the tasks of KT and AES which is positive. Moreover, we can see that the area of KT is extremely saturated with limited change in scope.

All these together show that there is a significant number of open problems ranging from replications studies in MPC to research on intermediate areas of AES and MPC which could be explored. At the same time, the study also shows that unlike other areas of research, EDM research has been mostly ad-hoc for specific studies like MPC and the datasets are not freely available for reworking and extending such problems, requiring a unified framework for the tasks.

## 2.5 Discussion

In this work, we surveyed three major tasks of EDM, namely MPC, KT, and Peer Feedbacks (AES & ASG). We reviewed 51 papers, got based on a systematic search of the three mentioned problems. During our survey, we found multiple dimensions work under each tackling application-centric, algorithmic and data specific challenges through improved annotations. We saw multiple open problems and challenges starting from a usage of simple data-driven techniques, the crisis of replication in case of MPC because of issues with datasets, lack of certainty in results from KT and limited cross-domain works in case of AES and ASG. We found consistency in results and evaluation metrics across the tasks of KT and AES which is positive. Also, we can see that the area of KT is extremely saturated with limited change in scope. All these together show that there is a significant number of open problems ranging from replications studies in MPC to research on intermediate areas of AES and MPC which could be explored. The study also shows that unlike other areas of research, EDM research has been mostly ad hoc for specific studies like MPC and the datasets are not freely available for reworking and extending such problems, requiring a unified framework for the tasks.

# 4 APPENDIX